\documentclass{amsart}

\usepackage{amssymb}

\usepackage[showframe=false, margin = 1.25 in]{geometry}
\usepackage[shortlabels]{enumitem}
\usepackage{tikz, xypic}
\usepackage{subfigure, float, booktabs, multirow}
\usepackage{comment}
\usepackage{booktabs}
\usepackage{hyperref}
\hypersetup{colorlinks,linkcolor={red},citecolor={olive},urlcolor={red}}
\usepackage{mathtools}
\mathtoolsset{showonlyrefs}
\usepackage[utf8]{inputenc}
\usepackage[english]{babel}

\definecolor{nicole1}{HTML}{009E73}
\definecolor{nicole2}{HTML}{E69F00}
\definecolor{nicole3}{HTML}{CC79A7}
\definecolor{nicole4}{HTML}{56B4E9}

\newcommand{\f}{\frac}

\newcommand{\D}{\mathcal D}
\renewcommand{\P}{\mathcal P}
\newcommand{\B}{\mathcal B}
\newcommand{\s}{\mathfrak s }
\renewcommand{\f}{\mathfrak f }
\newcommand{\m}{\mathfrak m }
\renewcommand{\c}{\mathfrak c }
\renewcommand{\L}{\mathcal L }
\newcommand{\h}{\mathfrak h }

\usepackage{theoremref}
\newtheorem{theorem}{Theorem}

\newtheorem{result}[theorem]{Result}

\theoremstyle{remark}

\theoremstyle{definition}

\title{Modeling COVID-19 spread in small colleges}

\author[Bahl]{Riti Bahl}
\address{Bard College; Mathematics}
\email{rb9956@bard.edu}

\author[Eikmeier]{Nicole Eikmeier}
\address{Grinnell College; Computer Science}
\email{eikmeier@grinnell.edu}

\author[Fraser]{Alexandra Fraser}
\address{Bard College}
\email{alexandragfraser@gmail.com}

\author[Junge]{Matthew Junge}
\address{Baruch College; Mathematics}
\email{Matthew.Junge@baruch.cuny.edu}

\author[Keesing]{Felicia Keesing}
\address{Bard College; Biology}
\email{keesing@bard.edu}

\author[Nakahata]{Kukai Nakahata}
\address{Baruch College; Mathematics}
\email{kukai.nakahata@baruchmail.cuny.edu}

\author[Wang]{Lily Z.\ Wang}
\address{Cornell University; Applied Mathematics}
\email{zw477@cornell.edu}

\thanks{This research was supported by NSF RAPID Grant \#2028892 and NSF Grant \#1953141.}

\begin{document}
\maketitle

\begin{abstract}
We develop an agent-based model on a network meant to capture features unique to COVID-19 spread through a small residential college. We find that a safe reopening requires strong policy from administrators combined with cautious behavior from students. Strong policy includes weekly screening tests with quick turnaround and halving the campus population. Cautious behavior from students means wearing facemasks, socializing less, and showing up for COVID-19 testing. 
We also find that comprehensive testing and facemasks are the most effective single interventions, building closures can lead to infection spikes in other areas depending on student behavior, and faster return of test results significantly reduces total infections.
\end{abstract}

\section{Introduction}

%As the Fall 2020 semester begins, u
Amid Fall 2020 of the COVID-19 pandemic, universities are rolling out a variety of interventions in hopes of safely offering in-person instruction  \cite{reopening}. Wrighton and Lawrence argued that ``best practices" should be followed, which include: testing, quarantine, contact tracing, facemask usage, and dedensification \cite{wrighton2020reopening}. While colleges in some parts of the world have successfully opened \cite{cheng2020safely}, the interventions that are about to be utilized in the United States are largely untested. A prominent example is the recent pivot by the University of North Carolina at Chapel Hill to remote instruction after an ``untenable" COVID-19 outbreak occurred during the first week of instruction \cite{unc}. Other major universities have subsequently followed suit in response to similar infection spikes upon reopening \cite{nd, ms}. 
%
%Colleges are unique in that administration policy, such as class scheduling and building access, can greatly impact the behavior of those on campus. 
Facing this uncertainty, simulation evidence may help inform policy and guide student behavior. 

Some models have already addressed COVID-19 spread on college campuses \cite{cornell1, cornell2, weeden2020small, university, colleges, durrett}. We discuss these in more detail in Section \ref{sec:others}, but note that their primary focus was medium-sized colleges.
%but note that none of these focused in deta on small colleges.
%but note that the primary focus was on colleges with over 20000 students. 
%
Given that there are more than 500 colleges in the United States with a student body of 4,000 or less
%, often centerpieces of smaller communities, 
that, in aggregate, serve over a million students, it seems important to specifically address this setting. We develop an agent-based model on a network to simulate COVID-19 spread through a small residential college.
%Agents move and pass infection through a network whose vertices represents physical spaces on campus. Reductions and simplifications are unavoidable. Nonetheless, t
The smaller population and campus allow us to make a relatively detailed model. 
%We build in features that capture unique aspects of life in a small college, which we subsequently use to measure the effectiveness of different interventions at mitigating disease spread. 
Beyond colleges, we believe that adaptations of our approach could be useful for modeling the effectiveness of interventions in other small,  closed-community residential settings such as military bases, single-industry towns, and retirement communities \cite{ho2014emerging, munanga2020critical}.

\subsection{Findings} \label{sec:findings}

The main statistics we consider are the total number of infections after 100 days on campus, and the basic reproduction number $R_0$, which is the average number of individuals directly infected by a single infected agent. Our model contains 2,000 students and 380 faculty. To standardize results, we start each trial with 0.05\% of students initially infected. %The only exception to this is when the campus population is reduced, we proportionally reduce the number of initial infections. 
More details on the specifics of our model can be found in Section \ref{sec:methods}.  
%computed by dividing the average number of agents directly infected by 10 on-campus students started in the exposed state by 10. 
Our main findings are given below and discussed further in Section~\ref{sec:strategies}.

\begin{figure}
    \centering
    \includegraphics[width = .7 \textwidth]
    {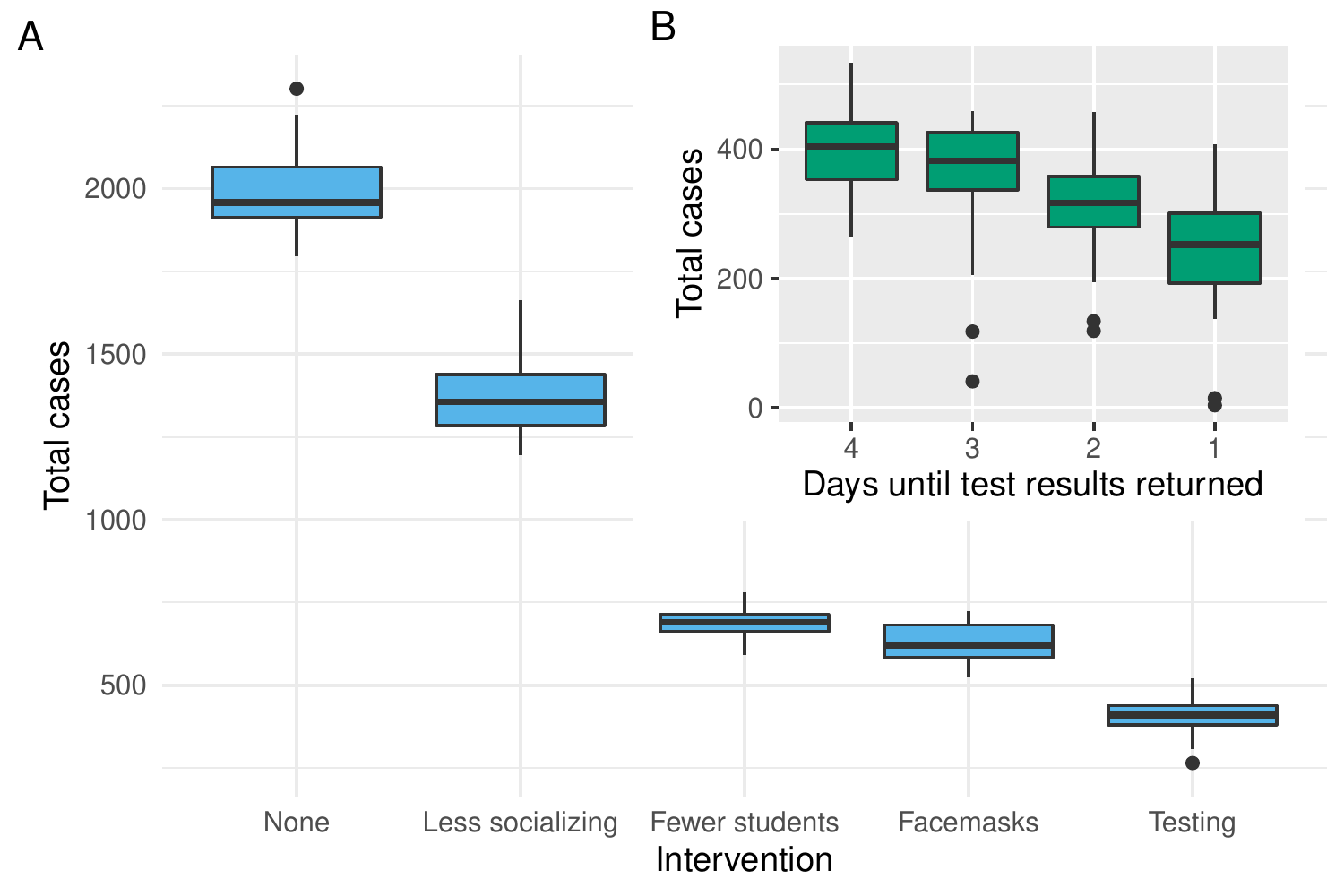}
    \caption{(A) The base model with single interventions applied. Note that the reduction in infections from ``fewer students"  is smaller than it appears since there are 50\% fewer people on campus in that intervention. (B) The impact of testing latency on a campus with $25\%$ fewer students and  testing and quarantine in effect.}
    \label{fig:marginals}
\end{figure}

% \begin{figure}
%     \centering
%     \includegraphics[width = .9 \textwidth]{curves}
%     \caption{Infection counts over time for the base model. The shaded region in Figure A contains two standard deviations.}
%     \label{fig:curves}
% \end{figure}

%\subsection{Summary of Results}
%Now for some details about the simulations and our main conclusions. 
%Our model contains 2380 agents: 1500 on-campus students, 500 off-campus students, and 380 faculty. The process 
 %We tune our base model with not interventions to have $R_0 = 3.9$. Our primary conclusions are the following:

\begin{figure}
    \centering
    \includegraphics[width = .6 \textwidth]{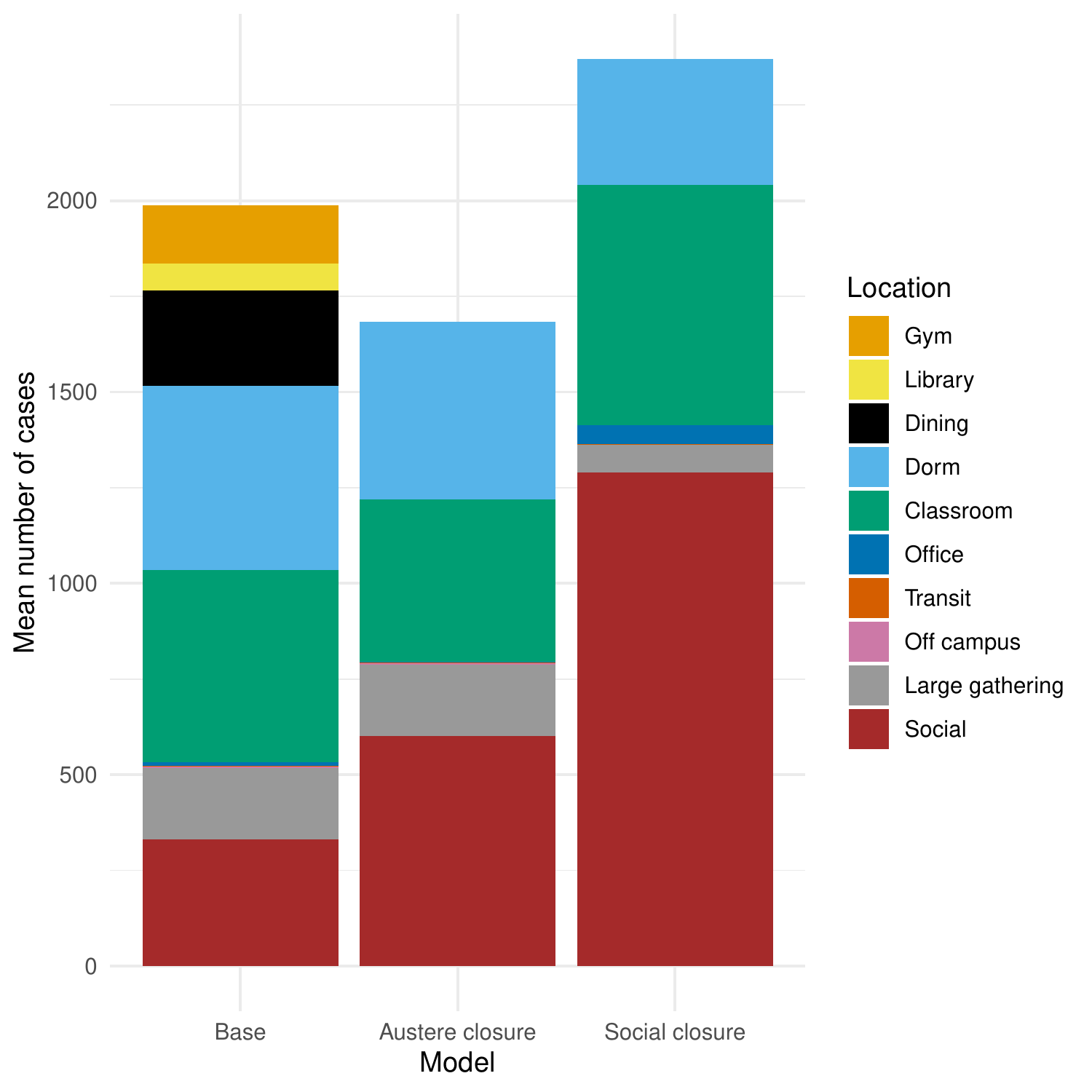}
    \caption{Total infections by room type in the base model and with the gym, library, and dining hall closed. In an ``austere closure", students spend any extra free time alone. In a ``social closure", students spend half of their free time socializing.}
    \label{fig:DH}
\end{figure}

\begin{result} \label{basemodel}
{Without intervention, most students and faculty become infected.}
\end{result}%(Without intervention there is a large epidemic)
With no interventions in place and $R_0$ set to $4.33$ (see Figure~\ref{fig:R0_marginals}), we found that the infection consistently reaches over 80\% of students and faculty (see Figure~\ref{fig:marginals} A). The most infections occur in classrooms, dorms, and while socializing (see Figure~\ref{fig:DH}). Peak infections occur between 40 and 50 days into the semester (see Figure~\ref{fig:curves}). 

\begin{result}\label{test+mask}
{Comprehensive testing and facemask compliance are the most effective single interventions.}  
\end{result}
%Sasha thinks "thorough" is better
%powerful used to be there
Weekly COVID-19 screening of 100\% of students with a two-day wait for test results
brings total infections from well above 1,900 to around 400 (see Figure \ref{fig:marginals} A). Alternatively, perfect facemask usage in public and social settings drops total infections to 630. 
%and $R_0$ to 2.3 with no other interventions in place. 

\begin{result}\thlabel{dininghall}
{Building closures may increase total infections.}
\end{result}
Closing the gym, library, and dining hall gives extra unstructured time to students. We find that if students are strict about passing that extra time alone, total infections decrease. However, if students spend half of that time socializing, we see a dramatic spike; nearly every agent in our model becomes infected (see Figure \ref{fig:DH}).

\begin{result}\label{latency}{Shortening time to receive test results reduces total infections.}  
\end{result}
We consider a campus at 75\% density with 50\% of students screened weekly for COVID-19 in addition to walk-in testing. No other interventions occur. We then vary the latency period to receive test results from four days down to one. Our model with a four-day latency period results in 400 total infections, compared to 250 with a one-day period (see Figure \ref{fig:marginals} B).

\begin{figure}
    \centering
    \includegraphics[width = .65 \textwidth]
    {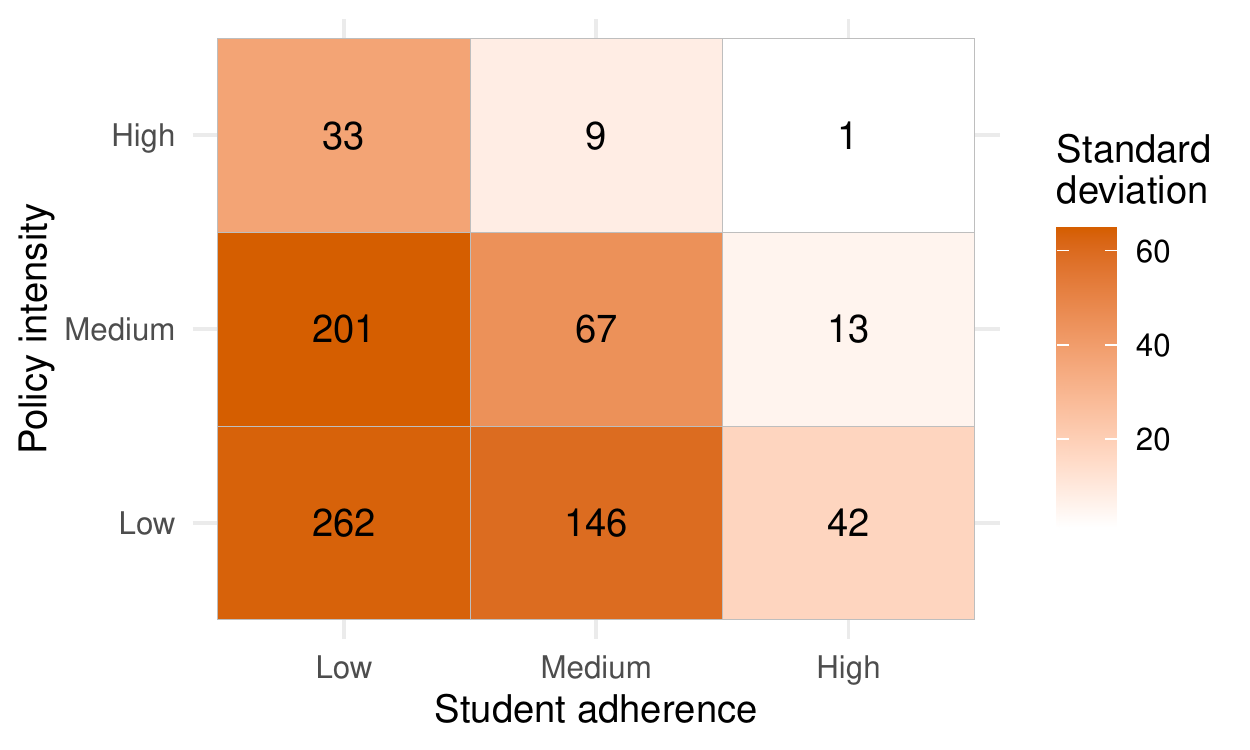}
    \caption{The total infection counts (numeric) and standard deviation (colorbar) for different policy and adherence intensities.}
    \label{fig:heatmap}
\end{figure}

\begin{result}\label{matrix}{Strong, unified administrative policy and student adherence result in the best outcomes.}
\end{result}
A novel part of our intervention design is that we separate student behavior from administrative policy. Specifically, students control facemask usage in social settings, compliance with screening tests, and time spent socializing. Administrators control the number of screening tests, testing latency, building closures, and the number of students allowed back to campus. 
We consider student adherence and administrative policy at low, medium, and high intensities. A high-intensity administrative policy by itself keeps total infections below 10 with moderate levels of student adherence. However, with less intense policy, we find that student adherence plays a crucial role. For example, total infections drop from 260 to 38 as student adherence increases with the low-intensity policy in effect. It is also worth noting that, under a high-intensity administrative policy, there is less variability as a result of student behavior.
See Figures~\ref{fig:heatmap} and \ref{fig:heatmap2} for more evidence and Table \ref{tbl:matrix} for a precise description of the interventions being applied. 

%\subsubsection*{Facemask compliance significantly reduces total infections} 

% \subsubsection*{Building closures displace infections to dorms and social spaces}
% %cause more
% A possibly unintended effect of closing libraries, gyms, and dining halls is that students will spend more time socializing and also in the common spaces of their dorms. One would hope that total infections go down, but these infections were instead displaced to other spaces. Students become more connected to the other inhabitants of their dorm building.

%and $R_0$ drops from 1.4 to 1.2.  

%\subsection{Limitations} 
%\subsection{Overview of model}

\subsection{Key Takeaways}

We outline some possible takeaways for administrators and students.

\subsubsection*{Administrators}
Our results suggest that strong administrative policy is needed, particularly regarding testing. Concerned administrators (and students) should check Table \ref{tbl:matrix} to see which intensity their reopening plan most aligns with. We emphasize that the low-intensity policy in our model tests 25\% of the student body weekly (Result \ref{matrix}). Without testing at or above this level, our results suggest that it will be hard to control COVID-19 spread. Test latency appears to make a difference as well; we advise that lowering the time to return results be a priority (Result \ref{latency}). Lastly, we demonstrate that building closures do not necessarily reduce total infections (Result \ref{dininghall}). Since social distancing can be more easily controlled in campus buildings, administrators may consider keeping buildings open. At the very least, students displaced by building closures should be encouraged to spend more time in isolation. 

\subsubsection*{Students}
  A serious and disciplined approach is needed from students (and administrators) to keep infections down (Result \ref{matrix}). We recommend that students wear facemasks in private settings, such as socializing, large gatherings, and common space in dorms (Result \ref{test+mask}). In light of the increased unstructured time resulting from building closures, it is especially important to spend more time alone rather than socializing (Result \ref{dininghall}). Given the impact of testing, students should cooperate fully with any required screening testing (Result \ref{test+mask}).

\subsection{Related work} \label{sec:others}
We know of five projects that specifically addressed COVID-19 spread on a college campus. Gressman and Peck \cite{university} used the University of Pennsylvania as a template to simulate different intervention strategies in an urban university with 22,500 students. This complemented recent work of Weeden and Cornwell \cite{weeden2020small} that studied how the degree of separation between students at Cornell University changes when some courses are switched to a remote or hybrid format. Around the same time \cite{university} was released, Frazier et al.\ posted a preprint \cite{cornell1} and, later, an addendum \cite{cornell2} that modeled how testing and quarantine could mitigate the spread of COVID-19 through Cornell's campus. Recently, Paltiel, Zheng, and Walensky studied the effectiveness of testing in a college with 5,000 students \cite{colleges}. Durrett et.\ al developed a mathematical model that rigorously demonstrated the benefits of limiting double occupancy dorms and of capping course enrollments \cite{durrett}.

To briefly summarize, \cite{weeden2020small} showed that a typical student directly interacts with about 4\% of the $22,000$ other students from common courses. However, the reach of a student jumps to 87\% when considering two degrees of separation, and to 98\% with three degrees. The authors further observed that removing large classes with an enrollment over 100 fails to disconnect the network and such interventions only increase the average graph distance between students by about $0.50$. For this reason, Weeden and Cornwell recommended taking further action than simply eliminating large courses. The authors also considered liberal arts colleges by restricting to the $4,500$ or so students in Cornell's College of Arts and Sciences. They observed that students in a liberal arts college are connected via short path lengths, but also through multiple paths. They inferred that this makes ripe social conditions for disease spread.

Frazier et al.\ also studied the Cornell student body, but rather than considering the network structure, they assumed a perfectly mixed population. They performed an SEIR model primarily taking into account the age of those infected, severity of symptoms, and amount of intervention through testing, quarantine, and contact tracing. They found that such interventions can suppress, but not completely contain the spread of COVID-19 during a semester. Despite fairly heavy intervention, asymptomatic spread results in 1,250 infections in their model. One shortcoming of their approach is that the perfect mixing assumption smooths over much of the structure inherent to a campus.

Paltiel, Zheng, and Walensky 
%zoomed in to the effectiveness and cost of screening strategies on a hypothetical medium-sized college with 5000 students. The study 
examined the epidemic outcomes and costs with varying test attributes and epidemic scenarios. They concluded that screening every two days with rapid, inexpensive tests results in a controlled number of infections with relatively low total cost. The authors acknowledged the logistical and financial challenges for university administrators even in the proposed testing scenario. The study did not consider other administrative strategies in combination with testing to restrict the spread of infection.  

Gressman and Peck 
%took advantage of the high degree of spatial and temporal structure present in college life to
built an agent-based model that incorporated more features of college life. Roughly speaking, on a given day in the model, an agent has approximately 20 contacts selected at random from different groups. These groups included residential, close academic, classroom contact, broad social, etc., and contact came with varying likelihoods of passing an infection. Their results suggested that large scale testing, contact tracing, and moving large classes online were the most impactful interventions. They further found that testing specificity is crucial for managing the number of people in quarantine. 
%The 
%Their approach %in \cite{university} 
%is tailored to the medium-sized urban university setting. 
%The way contacts occur is notably unlike most other agent-based models which typically have several orders of magnitude more agents interacting in a to-scale physical environment \cite{chang2020modelling, gomez2020infekta}. 
%Gressman and Peck
The authors observed that their model has limited applicability to small colleges \cite[p.\ 16]{university}. The important difference, in their view, is that students in a small college have fewer, but closer contacts compared to those at a large university. However, they pointed out that, without additional data, the different likelihood of infection may be a ``difficult feature to reasonably quantify or calibrate." 

%\todo{I feel that this paragraph should have a `the way our paper differs from these other studies is ...' vibe - NE}
%While we still lack descriptive data, o
One way we specifically account for social interactions is the introduction of ``social spaces'' into the network. Each student frequents two social spaces at which they contact a subset of roughly 20 other students. This generates two internally correlated, but externally independent friend groups.  More broadly, %we address this difficulty, 
we draw inspiration from larger agent-based models in which agents diffuse through a to-scale environment according to simple routines \cite{chang2020modelling, gomez2020infekta}. %The routines compounds into a complex system of interactions. 
We set the physical network and agent schedules as realistically as possible, then let the academic, residential, and social interactions tune to these choices. This philosophy distinguishes our approach from the models for COVID-19 spread in colleges mentioned above.

%More broadly, we make many simplifications, and, at times, our best judgement for parameters. Nonetheless,
%We believe that our paradigm offers a new, useful perspective on modeling COVID-19 spread.
%in a small college. 

% \subsection{Organization}

% Sections \ref{sec:space}, \ref{sec:agents}, and \ref{sec:infection} describe our model. %In particular, the space, agent behavior, and infection dynamics, respectively. 
% Section \ref{sec:interventions} describes individual interventions. Section \ref{sec:strategies} displays simulation results for different intervention strategies. 
%Appendix \ref{sec:appendix} has tables describing the parameters we use.

\section{Methods} \label{sec:methods}

In this section, we describe the network, agent behavior, and infection dynamics in our base model for a campus with no interventions in place. We conclude by describing different interventions. 

%\subsection{Overview of methods}
Buildings are star graphs whose cores represent shared spaces and leaves represent rooms or sections of the building (see Figure \ref{fig:campus}). 
%The cores are wired together via a single transit vertex, that represents the connective infrastructure between buildings. 
%The campus is utilized by 2380 students and faculty. 
%1500 on-campus students, 500 off-campus students as well as 380 faculty. 
Each agent is assigned a fixed schedule that determines their motion through the network which updates hourly (see Table \ref{tbl:schedule}). Infection dynamics follow an SEIR model (see \eqref{eq:SEIR}) where agents transition from the susceptible to the exposed state with probability proportional to the number of nearby infected agents scaled by the riskiness and size of the space (see \eqref{eq:pv}). 
%This captures unique features of colleges with 4000 or less students. 
We set the parameters (see the Appendix) to reflect the unique features of a small college campus---small classes; tightly knit, but diverse social groups; a primary dining hall, gym, and library---as well as our present understanding of the biology of COVID-19. We then overlay various interventions on the base model and measure their effectiveness.
%of different interventions at mitigating disease spread.  

\subsection{Space} \label{sec:space}

\begin{center}
\begin{figure}
\includegraphics[width = .7 \textwidth]{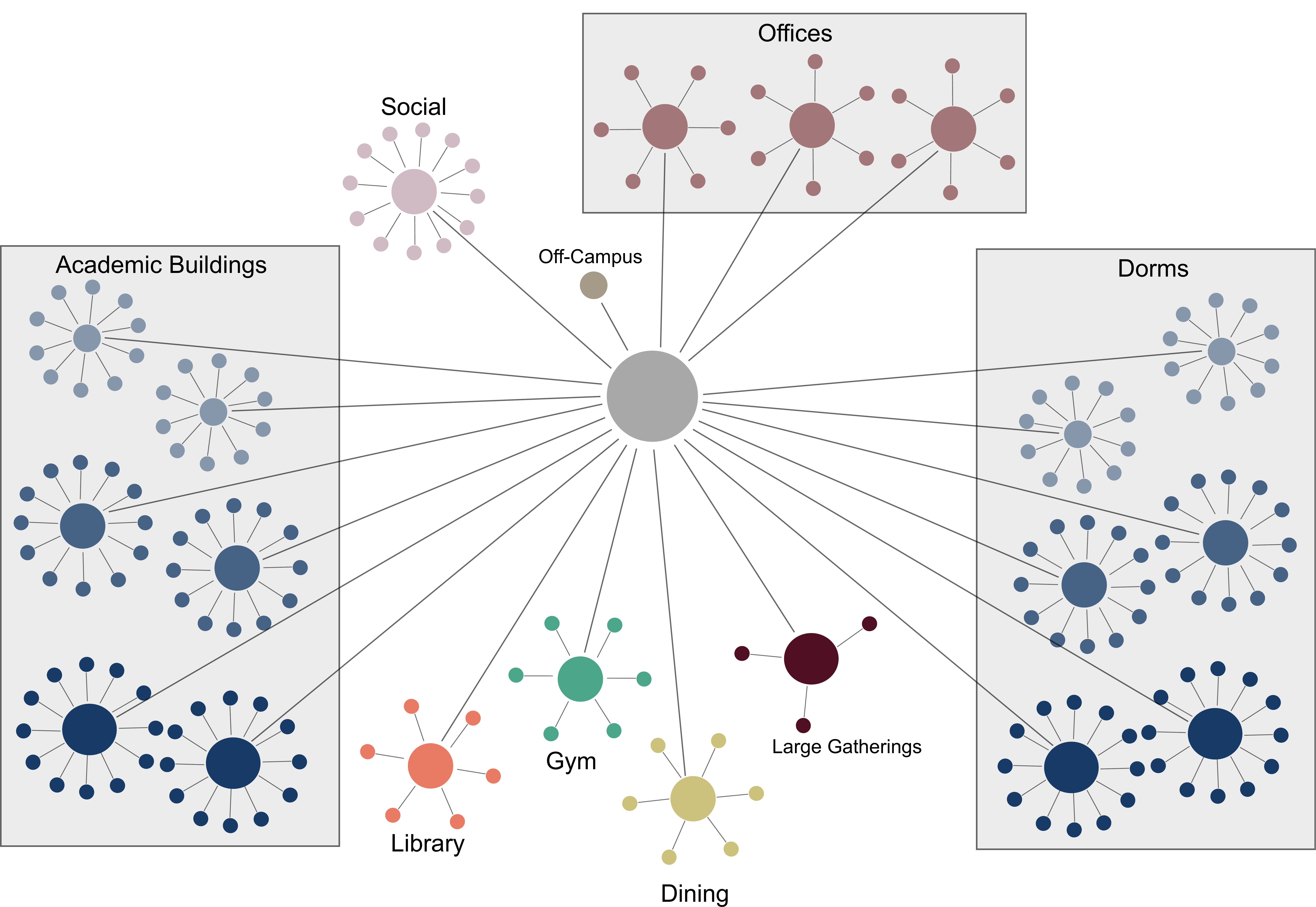}	
\caption{Schematic %(top) and implementation (bottom) 
of the network.} \label{fig:campus}
\end{figure}
\end{center}

Many of our decisions regarding our network draw inspiration from the campuses of Bard College and Grinnell College which exemplify small, relatively isolated, residential colleges. The basic building blocks are star graphs representing dorms, academic buildings, dining halls, gyms, social spaces, offices, and off-campus.  The core of each star represents shared space in the building such as hallways, bathrooms, lobbies, etc. The leaves represent either specific rooms or sections of the building. See Table \ref{tbl:buildings} for specifics. The core of each star connects to the transit vertex which represents the connective space between buildings. Note that the graph diameter is 4. See Figure \ref{fig:campus} for a schematic. %as well as a complete representation of the network.

%\subsection{Vertex Types}

\subsubsection*{Dorms, Classrooms, Academic Buildings} Are either small, medium, or large depending on the number of single and double rooms (Dorms), the number of seats (Classrooms), or the number of classroom sizes (Academic Buildings).
%	\item[Faculty Offices] There is one office building for the faculty in each division. Each faculty member is given a leaf. 
	%Each leaf represents either a single or double occupancy room. For simplicity, we assume that there are no triple rooms. There are three dorm sizes with the following ratios of room counts:  
%\begin{center}
%\begin{tabular}{c|cc}
%Dorm Building Size & Singles & Doubles \\
%\hline
%Small & 5 & 5 \\
%Medium & 15 & 15 \\
%Large & 25 & 25	
%\end{tabular}
%\end{center}
%Dorms that are significantly larger occupancy than 75 are divided into multiple \ttt{Large} dorms. For example, a dorm with occupancy 200 is divided into three \ttt{Large} dorms. 

%	\item[Classrooms] Are either small, medium, or large depending on the number of seats. 
%\begin{tabular}{c|cc}
%Classroom Size & Enrollment Capacity  & Room Capacity \\
%\hline
%Small & 10 & 15 \\
%Medium & 15 & 20 \\
%Large & 20 & 30	
%\end{tabular}
%\end{center}

%\item[Academic Buildings] Are either \ttt{Small}, \ttt{Medium}, \ttt{Large} and are designate to specific divisions among STEM, Humanities, and Arts.

%\begin{center}
%\begin{tabular}{c|ccc}
%Academic Building Size & Small & Medium & Large \\
%\hline
%Small & 3 & 0 &0  \\
%Medium & 2 & 3 &0 \\
%Large & 5 & 3& 3	
%\end{tabular}
%\end{center}

%\begin{center}
%\begin{tabular}{c|ccccc}
%Academic Building Size & Small & Medium & Large & Total Seats  & Proportion\\
%\hline
%Humanities & 2 & 2 &3  & 655 & .524\\
%STEM & 1 & 2 & 1 & 315	& .252	\\
%Arts & 2 & 1 & 1 & 280 & .224 
%\end{tabular}
%\end{center}

\subsubsection*{Dining Hall, Gym, Library, Faculty Offices} Are modeled by star graphs with six leaves. The leaves represent sections of the buildings. Our network has one gym, one library, one dining hall, and three faculty offices.
	%The core vertex represents shared space such as the entry, service areas, and bathrooms. The leaves represent tables customers eat at. Our model has a single dining hall with 100 leaves. 
%	\item[Libraries] Are modeled by star graphs in which the core represents shared spaces (bathrooms, hallways, entries) and the leafs represent areas students study at.
%	\item[Gyms] Gyms are represented by single vertices.
%	%The hub represents common spaces such as the entry, hallways, and locker rooms. The leafs represent specific areas of the gym such as weight and cardio rooms, pools, courts, etc. 
	\subsubsection*{Social Spaces} Are leaves of a star graph. The spaces represent social gatherings (study sessions, work groups, parties, casual social groups) that occur at various locations on campus. There are 100 such leaves. The core has no meaning, but is included for the sake of consistency in the underlying network.
	\subsubsection*{Transit Space} Is a single vertex that represents the paths, halls, and rooms that connect the other spaces.
%	\item[Private Space] Is a single vertex that represents time spent out of contact with others. For example, time spent during the day in a student's dorm room, time spent studying in a library alone, etc. 
	\subsubsection*{Off Campus} Is a single vertex that represents all space off campus.

\subsection{Agent Behavior}  \label{sec:agents}
In this section, we describe the types of agents, the way they are assigned schedules, and how they move through the network.

\subsubsection*{Agent types}

There are $n = 2,380$ total agents in the model; with $n_c =1,500$ on-campus students, $n_o= 500$ off-campus students, and $n_f = 380$ faculties. Agents are assigned a subtype that designates their division among STEM, Humanities, and Arts. We write $n_*^i$ with $i=1,2,3$ and $* \in \{c,o,f\}$ to denote the counts of STEM ($i=1$), Humanities ($i=2$), and Arts ($i=3$) agents.
We assume that STEM students are 50\% of the student body, Humanities students are 25\%, and Art students are 25\%. 
%i.e, $$n_o^1 + n_c^1 = n/2; \quad 	 n_o^2 + n_c^2 = n/2; \quad  n_o^3 + n_c^3 = n/4.$$
	Note that the division designations are interchangeable so these proportions represent whatever specialty a small college may have.

% 	\subsubsection*{On-Campus Student} Are identified by their dorm room. There are $n_c=1500$ many such students. On-Campus students randomly assigned a dorm room, starting with filling the singles. 
% 	\subsubsection*{Off-Campus Student} Begin and end each day off-campus. There are $n_o=500$ many such students. 
% 	\subsubsection*{Faculty} Begin and end each day off-campus. There are $n_f=380$ many faculty.

\subsubsection*{Agent Schedules}

%\subsubsection*{Time}
Days are classified as either $A$, $B$, $W$, or $S$. $A$ and $B$ days are distinguished by alternating class schedules. $W$ days represent weekends (Friday and Saturday) on which no instruction occurs and students socialize. To introduce some space into schedules, we  include Sundays ($S$) on which students either stay in their dorms or off-campus all day. A day is divided into 14 one-hour increments spanning from 8:00 -- 22:00 (the time $N$:00 will be abbreviated by $N$). Classes take place in two-hour increments starting at 10, 12, 14, and 16. 

%\subsubsection*{Class schedules}
We write each seat in a class on a given day and time as a $4$-tuple $(d,t,r,c)$ where $d \in\{A,B\}$, $t \in \{ 10,12,14,16\},$
$r$ is a classroom, and $c$ is a chair in $r$ (so $1\leq c\leq$ the enrollment capacity of room $r$). Let $\mathcal C$ be the set of all distinct seats $(d,t,r,c)$. Let $\mathcal C_1$ denote the set of all tuples whose building is designated a STEM building, and similarly for $\mathcal C_2$ and $\mathcal C_3$ for Humanities and Arts, respectively. Let $\mathcal C = \mathcal C_1 \cup \mathcal C_2 \cup \mathcal C_3$. To randomly assign classes, students with subtype $i$, one after the other, sample two elements uniformly at random from $\mathcal C_i$ and then two elements uniformly at random from $\mathcal C$ without replacement. If two selections conflict in time, classrooms are resampled until there are no conflicts. 

%\subsubsection*{Day schedules}
Once an agent obtains a class schedule, the remaining time slots are filled in according to the following rules. For each building in the schedule that is not a dorm or academic building, the agent is assigned to a uniformly sampled leaf, which they exclusively visit. The one exception concerns social spaces. For these, students are assigned a leaf for class days, and a leaf for the weekend. Since there are 100 social space leaves, on average 20 students are assigned to each leaf. Being assigned to two leaves makes it so agents interact with two social groups that are correlated within, but uncorrelated to other groups.	

For \emph{on-campus students}, each day begins and ends in their assigned dorm room at $8$ and $22$.  Each day type has one visit to the dining hall in the time slots 8--11, 12--15, 17--20. The afternoon slot 12--15 is skipped if the student has classes during that time. Lastly, each day type has a gym visit with probability $g$. The remaining slots are assigned to uniformly sampled social spaces with probability $s$, a library leaf with probability $\ell$, or the agent's assigned dorm room with probability $1-s - \ell$.

% Here is a sample schedule depicting a single class day for an on-campus student. See Table \ref{tbl:schedule} for a more detailed example.

% \begin{align}\begin{tabular}{l|lllllllllllllll}
% Hour &8 & 9  & 10    & 11    & 12 & 13 & 14    & 15    & 16 & 17 & 18 & 19 & 20 & 21 & 22 \\
% \hline
% Vertex Type &D & DH & $C_1$ & $C_1$ & DH & S  & $C_2$ & $C_2$ & DH & L  & D  & L  & S  & D  & D 
% \end{tabular} \label{eq:schedule}
% \end{align}

For \emph{off-campus students}, $A$ and $B$ days begin and end at the Off Campus vertex at times 8, 9 and 18--22. On $W$ and $S$ days the student remains at the Off Campus vertex all day. On $A$ and $B$ days, an off-campus student has one visit to the dining hall in the time slots 12--15, if the class schedule allows it. Each day type contains a gym visit with probability $g$ at a randomly chosen available time slot. The remaining slots are spent in a social space with probability $s$, at the library with probability $\ell$, and otherwise off-campus. 

For \emph{faculty,} $A$ and $B$ days begin and end with the agent at the Off Campus vertex at times 8, 9 and 18--22. On $W$ and $S$ days the faculty remains at the Off Campus vertex all day. If possible, the agent goes to the faculty leaf of the dining hall at a uniformly chosen time from 11--13.  The remaining slots are spent in the appropriate Division Office vertex.

\subsubsection*{Agent Paths}

Once an agent is assigned a schedule it remains to define the path the agent follows to move between each location. Suppose an agent is moving from a leaf of the core vertex $v$ to a leaf of the core vertex $u$. They do so by moving to $v$, to the transit vertex, to $u$, and then to the target leaf of $u$. We assume that transit occurs at the end of the hour and interacts with any other agents that move through the spaces $u,$ the transit vertex, and $v$ at the end of the same hour.

\subsection{Infection spread}  \label{sec:infection}

\subsubsection*{Agent States} \label{sec:states}
Agents are in states $S, E, I^{a}, I^{m}$, $I^{e}$ and $R$ corresponding to Susceptible, Exposed, Infected Asymptomatic, Infected Mildly Symptomatic, Infected Extremely Symptomatic, and Recovered. Agents transition through the states in the following manner:
%\begin{equation}\small \xymatrix{
%& & & \fbox{$I^{e}$} \ar[d] \\
%\fbox{$S$} \ar[r] & \fbox{$E$}\ar[r]  & \fbox{$I^{a}$} \ar[ru]^{e	} \ar[rd] \ar[r]^a  & \fbox{$R$} \\
%& &  & \fbox{$I^{m}$} \ar[u]}  \label{eq:SEIR}
%\end{equation}
%
\begin{equation} \xymatrix{
\fbox{$S$} \ar[r] & \fbox{$E$}\ar[r]  & \fbox{$I^{a}$} \ar[r] \ar@/^1pc/[rr]^{(1-a)e} \ar@/^3pc/[rrr]^a & \fbox{$I^{m}$} \ar@/_1pc/[rr] & \fbox{$I^{e}$} \ar[r]&  \fbox{$R$} }  \label{eq:SEIR}
\end{equation}

%\begin{align}\xymatrix{
%\fbox{$S$} \ar[r] & \fbox{$E$}\ar[rd]  & \fbox{$I^s$} \ar[r]  &\fbox{$R$} \\
%& & \fbox{$I^a$} \ar[u] \ar[ru]^a &} \label{eq:SEIR}
%\end{align}
 We let $I_v^a(d,t)$ denote the number of agents in state $I^a$ at site $v$ at time $(d,t)$ and similarly for the other states. Describing how and when agents transition from state $S$ to state $E$ is the subject of the next section. The other transitions are simple to describe:
 \begin{itemize}
 	\item Agents stay in state $E$ for $T_E=2$ days. After which, they transition to state $I^a$.
 	\item Each agent in state $I^a$ transitions to state $R$ after  $T_{I^a}=10$ days (from the day of infection) with probability $a$. Otherwise, after $T_{I^a}^*=2$ days the agent transitions to state $I^e$ with probability $e$ and to state $I^m$ with probability $1-(a+e)$. 
 	\item Each agent in state $I^e$ transitions to state $R$ after $T_{I^e}=10$ days. However, after $T_{I^e}^* = 5$ days the agent spends the subsequent time in their dorm room. This represents a student becoming ``bed-ridden," i.e., too sick to leave their room.
 	\item Agents in state $I^m$ transition to state $R$ after $T_{I^m}=10$ days.
 \end{itemize}

%\section{Viral dynamics}

\subsubsection*{The base probability of infection}
The vertex $v$ at time $(d,t)$ has \emph{infection probability} 
\begin{align}p_v(d,t) = r_v\frac{I_v^e(d,t)+I_v^m(d,t) + 0.5 I_v^a(d,t)}{C_v} p \label{eq:pv}.\end{align}
The parameter $C_v$ is the \emph{capacity} of $v$ and $r_v \in \{0,1,2,3\}$ is the \emph{risk multiplier} for infection spread in that space. Each of the $S_t(v)$ susceptible agents at $v$ at time $t$ independently enters state $E$ with the probability at \eqref{eq:pv}. Note that we set the infectiousness of an agent in state $I^a$ to half that of an agent in the other infected states \cite{facemask}. %
 The constant $p$ is the \emph{tuning parameter} that allows us to control global infectiousness.
 
 %at all sites. It is generated by running the simulation with  ten on-campus students in the exposed state. All other agents can become exposed, but we do not allow them to advance through the other states. If $x$ people become infected, then we set $p$ so that $(x-10)/10$ is the desired value of $R_0$. 

\subsubsection*{The risk and capacity parameters}

The parameter $r_v$ is chosen based on time spent, the proximity of agents in the space, and the typical amount of respiration---i.e.\ time spent talking aloud or exercising---in a given space. For example, $r_v$ is higher in the gym compared to the library. 
We set $C_v$ equal to ten times the core capacity for buildings with known capacities in advance (dorms and instructional buildings). The factor of ten is to dilute the number of people in the core at a given time (otherwise all of the agents would simultaneously be in that location). Ten is chosen since a passing time between classes is about that duration in minutes. 
The capacities for the dining hall, library, gym, and social spaces are set empirically to match the typical occupancy of the building.  See Table \ref{tbl:capacaties} for all of the $C_v$ and $r_v$ values.
%We set each leaf to be 1/6 the capacity of the core. The diluting factor of 10 is not used for cores since the common spaces for these buildings are used more frequently than in dorms and instructional buildings.
%Other buildings---dining hall, library, gym, and offices---we set $C_v$ through empirical observation. 
%Specifically, $C_v$ is set at twenty times the maximum number of agents ever at each core and leaf over a simulated week on the campus. The transit hub we dilute even further to have capacity $n^2$ since this occurs in a presumably open, large space. 
%For social space leaves, we set $C_v(d,t)$ dynamically to either $5\lceil x/5 \rceil$ with $x$ the number of agents at $v$ at time $(d,t)$. 
%equal the current number of agents at that space at that time. This models the idea that proximity in a private social interaction is independent of the venue.

\subsubsection*{Exceptions}
Two exceptional spaces, where the infection dynamics are not exclusively governed by \eqref{eq:pv}, are off-campus and large gatherings.
%\subsubsection*{Off-campus} 
Upon leaving the off-campus vertex at $t=8$, each agent in state $S$ transitions to state $E$ with probability $o$. For agents \emph{returning from off-campus}, we choose $o= .125/ (n_o + n_f)$ so that, on average, one off campus agent becomes infected every 8 class days (two weeks).
  %
%\subsubsection*{Large gatherings}  \label{sec:LG}
For \emph{large gatherings}, half of the student agents (both on- and off-campus) are denoted as ``social." We simulate large informal gatherings (e.g., parties or organized social events) by drawing three random subsets $G_1, G_2, G_3$ of agents designated as social at the end of each week. Each $G_i$ has size uniformly and independently sampled from $[20,60]$. The $G_i$ are sampled independently and are not necessarily disjoint. Each susceptible agent at a large gathering becomes infected according to \eqref{eq:pv} with $r_v = 3$ and $C_v= 40\lceil |G_i| / 40\rceil,$ i.e., $C_v =40$ if $|G_i| \leq 40$, and $C_v = 80$ if $|G_i| > 40$.

\subsection{Types of intervention}  \label{sec:interventions}
We consider a variety of interventions that broadly include: facemasks, testing/quarantine, building closures, less socializing, and dedensification, which we describe in more detail below.

\subsubsection*{Facemasks} We assume that agents never wear facemasks at dorm and dining hall leaves. There is partial compliance at dorm cores, social space leaves, and large gatherings. All other vertices have perfect compliance. Let $\f\in \{0.50, 1\}$ be the proportion of compliant agents. We implement this intervention by randomly selecting the corresponding percentage of agents who always wear a facemask at partial compliance vertices. We assume that wearing a mask reduces an agent's infectivity by a factor of $\m=0.5$ (which is the conservative estimate from \cite{facemask}). So, an infected agent wearing a mask is a factor of $\m$ less infectious, and a susceptible agent wearing a mask is a factor of $\m'=0.75$ times the probability of becoming infected at each time location. That facemasks protect the wearer is supported by recent evidence from \cite{mask2}. For example, a susceptible person wearing a mask in room $v$ at time $(d,t)$ will become infected with probability
\begin{align}p'_v(d,t) =  \m' \frac{\m M_v(d,t) +I_v(t,d)}{C_v} p\label{eq:mask}\end{align}
rather than \eqref{eq:pv}, where $M_v(d,t)= M^e_v(d,t)+M^m_v(d,t) + 0.50 M^a_v(d,t)$ are the number of agents in the infected state wearing a mask at $v$ at time $(d,t)$ and $I_v(d,t) = I^e_v(t)+I^m_v(d,t) + 0.50  I^a_v(d,t)$ are (weighted by infectiousness) number of infected agents in the infected state not wearing a mask at $v$ at time $(d,t)$.

\subsubsection*{Testing and Quarantine} 

In line with \cite{university}, we assume a false positive rate of $FP =0.001$ for agents tested while in the susceptible or exposed state, and a false negative rate of $FN = 0.03$ for agents tested while in an infected state.

\textbf{Screening:} We assume that $ \P \in \{ 0.25, 0.50, 1\}$ of the student body is screened per week.
%screening tests are performed per day that classes are held. Thus, 400, 1000, or 2000 tests are performed per week. O
Only students are screened, and the screening is applied throughout the entire student body on a repeating cycle. The \emph{latency period} $\L \in \{1, 2, 3, 4\}$ is the number of days to receive results. After the latency period, the infected agents from the batch who test positive are placed in the quarantine state for 14 days, after which they transition to the recovered or susceptible state depending on whether or not the test was correct.
We consider $\c \in \{ 0.80, 0.90, 1\}$ the level of compliance for agents in state $I^a$ to get screened. This means that each time an agent in the $S, E,$ or $I^a$ state is selected for screening, the agent skips taking the test with probability $1-\c$.

\textbf{Walk-ins:} For each day following the first that an agent enters state $I^e$ or $I^m$, that agent opts to be tested with probabilities $q_e=0.95$ and $q_m = 0.70$. After this, the agent enters the quarantine state with probability $1-FN$ depending on if they are in state $I^m, I^e$, or $I^a$. For example, the probability an agent in state $I^e$ enters the quarantined state $k$ days after entering state $I^e$ is $(1-FN)(1-q_e)^{k-1}q_e$. The probability $q_*$ represents an agent ignoring symptoms on a given day and waiting to take the test. We assume that walk-ins immediately begin quarantine, but re-enter the campus if they receive a false negative result.

\subsubsection*{Closures} We assume that buildings in $\mathcal B \subseteq \{L, G, DH, O, LG\}$ are closed. If the library ($L$), gym ($G$), or dining hall ($DH$) are closed, time spent at the space is replaced in a student's schedule with time in the student's dorm room or off-campus, depending on the type of student, with probability $\h \in \{0.50, 0.75,  1\}$. Otherwise, the agent goes to the social space. When facing a building closure, faculties spend that time in their office instead. 
%
%When the dining hall ($DH$) is closed, the core remains open, but no infection occurs at the leaves. This models a ``self-serve, no dining-in" policy. 
When faculty offices $(O)$ are closed, no infection occurs there, and we assume faculty only spend time in the classes they teach. When large gatherings ($LG$) are removed, we turn off the large gathering component.

\subsubsection*{Dedensification} For \emph{medium dedensification} we remove $\D=650$ agents: 250 on-campus, and 250 off-campus students, as well as 150 faculty at random. For \emph{high dedensification} we remove 1300 agents:  500 on-campus students, 500 off-campus students, and 300 faculty from the campus. The first students to be removed are those in double rooms. %The 500 for $\D=1300$ is chosen so that, in our model, all dorm rooms are singles.  

A few technicalities emerge with dedensification in effect. Courses in either degree of dedensification are assumed to be hybrid. All classes continue to meet, but the removed students attend class remotely. We assume that large gatherings do not occur whenever dedensification is in place. 
%Note that for simplicity in our implementation we have screening testing apply to the entire population of 2000 students (even those who are not present). Still, with high dedensification and screening with $\mathcal P = 1$ about half of the enrolled students are included in eatch batch. Even with this randomness, all but a few students will be screened weekly.  
Lastly, a dedensified campus will naturally have fewer initially infected agents. We account for this by starting with $\mathfrak i \in \{5, 7, 10\}$ on-campus students infected, with $\mathfrak i$ chosen to be approximately $0.05\%$ of the students and faculty still utilizing the campus. When $\D=650$, we assume that $\mathfrak i = 7$, and when $\D=1300$ we assume that $\mathfrak i=5$.

\subsubsection*{Less socializing} We replace time in social spaces with time spent at the student's dorm room or the off-campus vertex depending on the type of student. This replacement is done to each occurrence of social space in an agent's schedule with probability $\mathfrak s \in \{ 0, 0.25, 0.75 \}$.

{%\small
\begin{table}
\begin{tabular}{l  l l l}
%\hline

\textbf{Parameter} & \textbf{Value} & \textbf{Description} & \textbf{Ref} \\
%\hline
\toprule
\textbf{Base Model} & & & \\
$(n_c; n_c^1, n_c^2, n_c^3)$ 	&	(1500; 750, 375, 375) & on-campus student counts by division & \cite{student_discipline_distribution}	\\
 $(n_o; n_o^1, n_o^2, n_o^3)$ & (500; 250, 125, 125) & off-campus student counts by division & \cite{student_discipline_distribution}  \\
 $(n_f; n_f^1, n_f^2, n_f^3)$ & (380; 190, 95, 95) & faculty counts by division & \cite{student_discipline_distribution, number_of_faculty} \\
 $(g,s,\ell)$ 	&	(0.15, 0.15, 0.15) & gym, social, and library probabilities  &\cite{hardy2011study, eagan2014american}\\
% $(g',s',\ell')$ 	&	(0.15, 0.10, 0.25)& off-campus: free-time probabilities &\\
 $o$ 	&	$0.125/(n_o+n_f)$ & off-campus infection probability &	\\
 $T_E$ 	&	2 & days in the exposed state &	\cite{cdc_2020_exposed}\\
 $a$ 	&	0.15 & probability of remaining asymptomatic &	\cite{mizumoto2020estimating}\\
 $e$ & 0.50  & probability of $I^a \to I^e$ & \cite{liguoro2020sars}\\
 $T_{I^a}$ 	&	10 & days in $I^a$ if asymptomatic & \\
$T_{I^a}^*$ 	&	2 & days in $I^a$ if symptomatic & \cite{wei2020presymptomatic}\\
 $T_{I^e}$ 	&	10 & days in $I^e$ if never bid-ridden & \cite{sanche2020early}\\
 $T_{I^e}^*$ & 5  & days in $I^e$ if bed-ridden & \cite{sanche2020early}\\
 $T_{I^m}$ 	&	10 & days in $I^m$  & \cite{world2020report}\\
 $R_0$ & 4.33 & average \# of new cases from 10 exposed & \cite{cao2020estimating}\\
 $p$ & $1.25$ & tuning parameter & \\
 $FP$ &  0.001 & false positive rate & \cite{university}\\
 $FN$ &  0.03  & false negative rate & \cite{university}\\
 %\hline
 \midrule
\textbf{Interventions} & & & \\	
$\f$ 	&	0, 0.50, 1 & facemask compliance &	\\
$\m$ & 0.50 & facemask reduced infectiousness & \cite{facemask} \\
$\m'$ & 0.75 & facemask protection from infection & \cite{mask2} \\
$\P$ & 0.20, 0.50, 1 & weekly percentage of students screened & \\
%$N$ & 100, 250, 500 & number of daily tests & \\
$\L$ & 1, 2, 3, 4 & latency period to receive results & \\
$\c$ & 0.80, 0.90, 1 & asymptomatic screening compliance\\
$q_e$ & 0.95 & probability of symptomatic walk-in test & \\
$q_m$ & 0.70 & probability of mild walk-in test & \\
$FP$ & 0.001 & false positive rate & \cite{university} \\
$FN$ & 0.03 & false negative rate & \cite{university} \\
$\mathcal B$ & $L, G, DH, O, LG$ & building closures & \cite{cdc_2020_considerations}\\
$\h$ & 0.50, 0.75, 1 & prob.\ of dorm/off-campus from bldg.\ closure &  \\
$\mathcal D$ & 0, 650, 1280 & dedensification amount & \cite{pisacreta_banes_2020}\\
$\mathfrak s$ & 0, 0.25, 0.75 & reduction in socializing & \\
$\mathfrak i
$ & 5, 7, 10 & initial infected cases with dedensification &\\
\bottomrule
\end{tabular}
%\vspace{ .25 cm} 
\caption{Parameters.} \label{tbl:parameters}
\end{table}
}

\section{Results} 
\label{sec:strategies}

There are over a hundred thousand distinct combinations of the five single interventions from Section \ref{sec:interventions}. Therefore, some care is required to decide what combinations provide useful insights. To this end, we reduce down to 20 strategies. We consider the following two statistics:
\begin{itemize}
	\item \emph{Total infections} are the total number of agents ever in the exposed state after running the model for 100 days with $\mathfrak i$ on-campus students initially in the exposed state. The value of $\mathfrak i \in \{5, 7, 10\}$ depends on the amount of dedensification in place and is not counted towards total infections. We perform 40 independent simulation trials for each model (with new schedules in each trial). Each trial takes a little over a minute to simulate on a home computer. It takes about a day on a single machine to run all 20 interventions 40 times each.

	\item $R_0$ is the average of the value $(x-10)/10$ over 100 simulations where $x$ is the number of first generation infections caused by ten randomly selected on-campus students initially in the exposed state with all other agents in the susceptible state. Unlike with total infections, the number of initially infected agents in the $R_0$ calculation (ten) does not vary with the level of dedensification. 

\end{itemize}

There are four different types of experiments.

\subsubsection*{Marginals} \label{sec:marginals} We apply single interventions at high-intensity to the base model. Specifically, we consider: no intervention, facemasks with $\f = 1$, high dedensification with $\D = 1300$, less socializing with $\s = 0.75$, and testing with $\mathcal P = 1$. The results are shown in Figures \ref{fig:marginals}, \ref{fig:R0_marginals}, \ref{fig:curves}, and \ref{fig:DH}.

\subsubsection*{Building closures} \label{sec:closures} We close the gym, libarary, and dining hall with $\h=0.50$ and $\h = 1$. No other interventions are applied. See Figure \ref{fig:DH}. 

\subsubsection*{Test latency} \label{sec:latency}  We fix the base model with medium dedensification ($\D = 650$) and testing with $\P = 0.50$. This means that there are about 25\% fewer students on campus, of whom 50\% are screened weekly. We then consider latency $\L \in \{1,2,3,4\}$. The results are shown in Figures \ref{fig:marginals} and \ref{fig:R0_marginals}.

\subsubsection*{Policy and Adherence} \label{sec:matrix} 
To address the problem of choosing which interventions to run among the many we could apply, we classify the single interventions as either an administrative policy, or a student adherence behavior.
 We group interventions by type and set each to one of three different intensity levels. This gives nine combined strategies, which we hope offer a practical perspective for students and administrators attempting to manage the risk of COVID-19 spread. The specific parameters used for low, medium, and high-intensity policy/adherence are given in Table \ref{tbl:matrix}. The results are shown in Figures \ref{fig:heatmap}, \ref{fig:heatmap2}, and \ref{fig:R0_matrix}. 

%A summary of our results is shown in Figure \ref{fig:heatmap}.

\begin{table}
\begin{tabular}{l@{\hskip 0.25in} l l l l@{\hskip 0.4in} l l l l }
& \multicolumn{4}{l}{\hspace{.5in}\textbf{Policy}}  & \multicolumn{4}{l}{\hspace{.2in}\textbf{Adherence}}\\
& $\P$ & $\L$& $\mathcal D$  & $ \B$ & $\f$ & $\c$  & $\h$ & $\s$    \\
%\cmidrule(lr){2-5} \cmidrule(lr){6-9}
\cmidrule[\heavyrulewidth]{1-9}
 Low & 0.25 & 4 & 0 & $\{G, L\}$ & 0 & 0.80 &  0.50 & 0  \\
Medium & 0.50 & 3 & 650 & $\{G, L, DH, LG\}$  & 0.50 & 0.90 & 0.75 & 0.25   \\
High & 0.75  &  2 & 1300& $\{G, L, DH, O, LG\}$   &1 & 1 & 1 & 1 	 	\\
\cmidrule[\heavyrulewidth]{1-9}
\end{tabular}
%\vspace{.15 cm}
\caption{The intervention parameter choices corresponding to different intensities for  administrative policy (left) and student adherence (right). 
} \label{tbl:matrix}
\end{table}

Recall, that our primary findings are:

\begin{enumerate}[label = \arabic*.]
    \item Without intervention, most students and faculty become infected.
    \item Comprehensive testing and facemask compliance are the most effective single interventions.
    \item Building closures may increase total infections.
    \item Shortening time to receive test results reduces total infections.
    \item Strong, unified administrative policy and student adherence result in the best outcomes.
\end{enumerate}
We now explain how these experiments support these results.

%The corresponding simulations support Results \ref{basemodel}, \ref{test+mask}, and \ref{dininghall}.

    \subsection*{Result \ref{basemodel}} In our base model, we set the tuning parameter $p=1.25$ so that $R_0=4.33$ (see Figure \ref{fig:R0_marginals} A).  This consistently leads to a large infection that reaches on average 1988 agents (see Figure \ref{fig:marginals}). Figure \ref{fig:curves} displays the evolution of the infection over time. The peak typically occurs between 40 and 50 days into the semester. Figure \ref{fig:curves} A shows two standard deviations of data. The breakdown of infection counts by building type are given in Figure \ref{fig:DH}. Dorms, classrooms, social spaces, and the dining hall make up the majority of cases. Large gatherings and the gym are next. 
    
    \subsection*{Result \ref{test+mask}} Figure \ref{fig:marginals} shows how weekly testing of 100\% of students with latency at $\L =2$, consistently reduces infections below 500. The value of $R_0$ is $1.28$ with a standard deviation of 0.34 (Figure \ref{fig:R0_marginals}). 
    The second most effective intervention is facemask usage. Total infections stay around 600-800 (Figure \ref{fig:marginals}) and the average value of $R_0$ is 2.42 with a standard deviation of 0.57 (Figure \ref{fig:R0_marginals}). Note that it is somewhat misleading how Figure \ref{fig:marginals} suggests high dedensification is comparably effective to facemask usage. This is because there are only half as many agents present during that intervention.  
    
    \subsection*{Result \ref{dininghall}} Figure \ref{fig:DH} shows the vertices where infections occur in the base model alongside the effects of closing the gym, library, and dining hall. With closures, we consider the settings with $\h = 1$ and $\h=0.50$. We call the case $\h=1$ an ``austere closure" since students are electing to pass the time slots they would have been in a closed building at either their dorm room or off-campus. With an austere closure, total infections drop from nearly 2000 to around 1700. The total number of infections in social spaces increases, since these  infections would normally occur earlier in a closed building, but instead occur later in a social space. The case $\h=0.50$ is a ``social closure" in which students go to social spaces with probability 0.50. The last column of Figure~\ref{fig:DH} shows a significant increase in infections. A huge increase in social space infections allows the infection to proliferate. We note that the final counts are unrealistic, since it seems unlikely to us that a college would remain open after so many students are infected. Nonetheless, the mixed effect of closing buildings is illustrated by these counts. 
    
    \subsection*{Result \ref{latency}} As $\L$ goes from $4$ to $1$ total infection counts drop from 403 on average to 247 and $R_0$ drops from 1.28 to 1.07. See Figures \ref{fig:marginals} B and \ref{fig:R0_marginals} B. One interesting feature is that the variance increases as $\L$ decreases. When $\L = 4$, the standard deviation in total infections is 60; but when $\L = 1$, the standard deviation is 87. The reason for the greater volatility is that shorter latency sometimes is very effective and completely controls the infection, and sometimes the infection spreads more quickly than testing can control, resulting in many infections (relative to the mean).
    
    \subsection*{Result \ref{matrix}} Figure \ref{fig:heatmap} shows that the average number of total infections drops from 262 to 1 as policy and adherence are strengthened. The standard deviation drops significantly as well. We see that total infections are reasonably controlled by high-intensity policy (top row of Figure \ref{fig:heatmap}). Notice that the value of $R_0$ is consistently below $0.50$ whenever the high-intensity policy is in effect (Figure \ref{fig:R0_matrix}). However, with low student adherence and high-intensity administrative policy, there are simulations with $R_0 >1$. On the other hand, if policy and adherence are both low then $R_0$ is on average around 1.25 and at times above $2$. Figure \ref{fig:heatmap2} displays the coefficient of variation (standard deviation/mean). The figure illustrates how low-intensity policy coupled with low adherence, even after normalizing for the mean, has the highest variation. Additionally, Figure \ref{fig:heatmap2} shows that high-intensity administrative policy can temper variation stemming from different levels of student adherence.

\subsection*{Limitations}
%We strongly encourage the reader to understand the inner-workings of our model, because all of our claims are restricted to that universe. 
The response to COVID-19 is unprecedented and untested. Data is limited. We were often forced to use our best guess for parameters (see Table \ref{tbl:parameters}). An unfortunate feature of our model is that it is sensitive to these choices; scaling many of our parameters by a small constant (say 2), can increase, decrease, and shift infection counts in the resulting outcomes. For example, doubling the capacity for social spaces from 10 to 20, shifts infections elsewhere. For this reason, we stress that all of our results should be strictly viewed through the lens of our model. The values we present for total infections and $R_0$ are meant to be illustrative of trends, rather than precisely forecasting what would happen. 
%Even if precise forecasting is limited, the reductions we observe still allow us to make inferences about the effectiveness of various interventions. 
%
Another limitation of our model is that the way infections occur makes contact tracing impractical to implement. Unlike \cite{university}, in which contacts are known, we assume perfect mixing on the level of rooms, so it is not possible to infer who did the infecting. Staff and visitors to campus are another noteworthy feature that our model is missing. In the future, we plan to add more variety in agent types and behavior and also consider other  interventions as well as combined strategies.

%\section{Conclusion}

%\vfill

\newpage 

\appendix

%\section{Supplementary Figures}

\section{Simulation results}

\begin{figure}[H]
    \centering
    \includegraphics[width = .8 \textwidth]{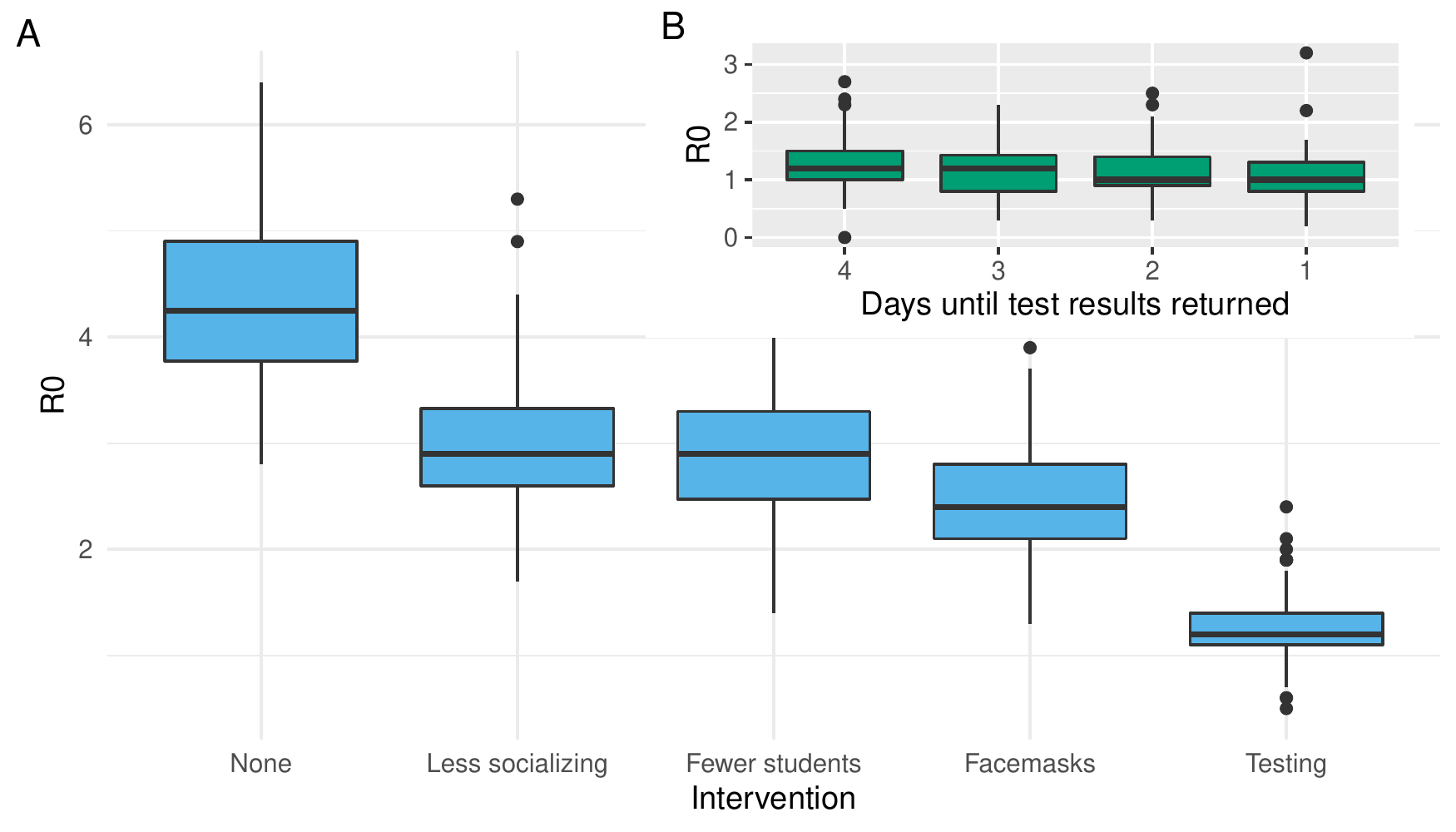}
    \caption{The values of $R_0$ for marginal interventions (A) and for testing latency (B).}
    \label{fig:R0_marginals}
\end{figure}

\begin{figure}[H]
    \centering
    \includegraphics[width=.9\textwidth]{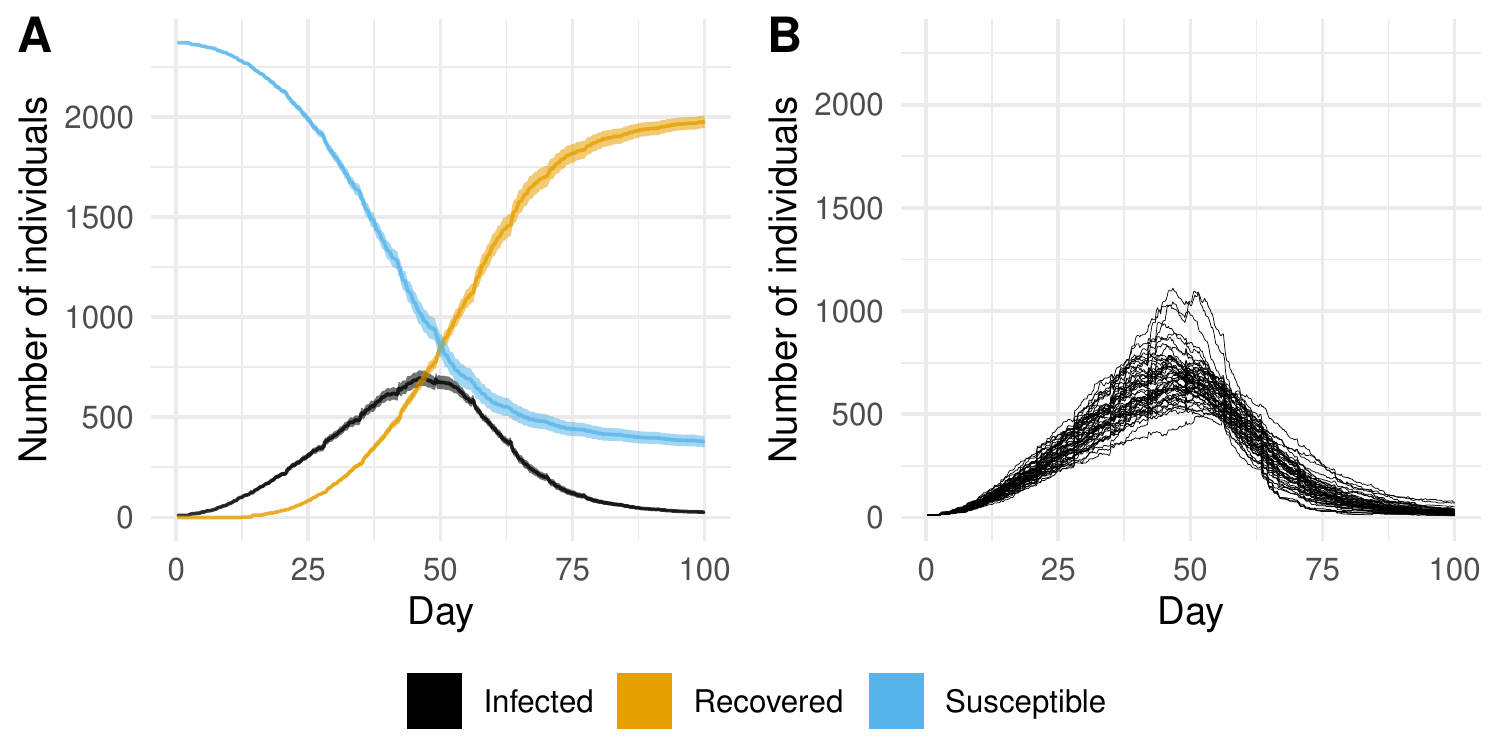}
    \caption{Agent states over 100 days in the base model. Panel A shows a 95\% confidence around the mean behavior from 40 trials. Panel B shows the infection counts for each trial.}
    \label{fig:curves}
\end{figure}

% \section{Simulations from Section \ref{sec:matrix}} \label{apx:matrix}

\begin{figure}[H]
    \centering
    \includegraphics[width = .7 \textwidth]{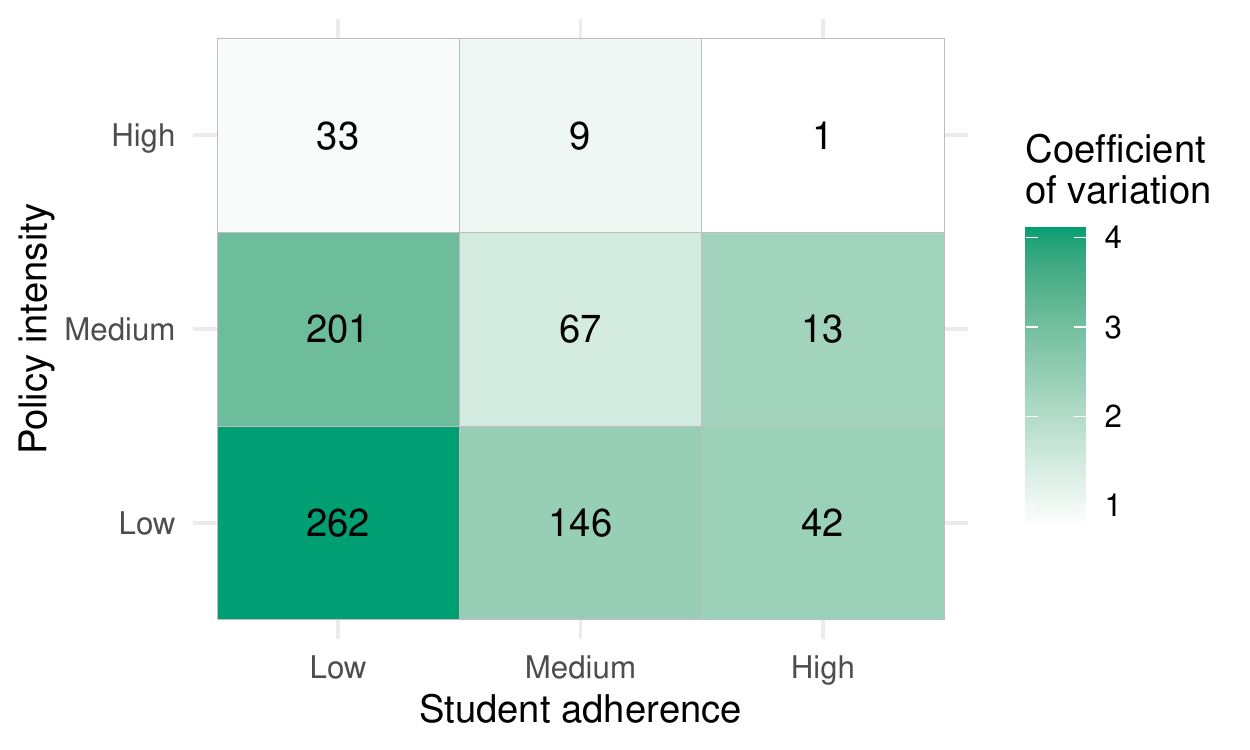}
    \caption{The total number of cases (numeric) and the coefficient of variation (standard deviation/mean; colorbars) for different policy and adherence intensity levels.}
    \label{fig:heatmap2}
\end{figure}

\begin{figure}[H]
    \centering
    \includegraphics[width = .8 \textwidth]{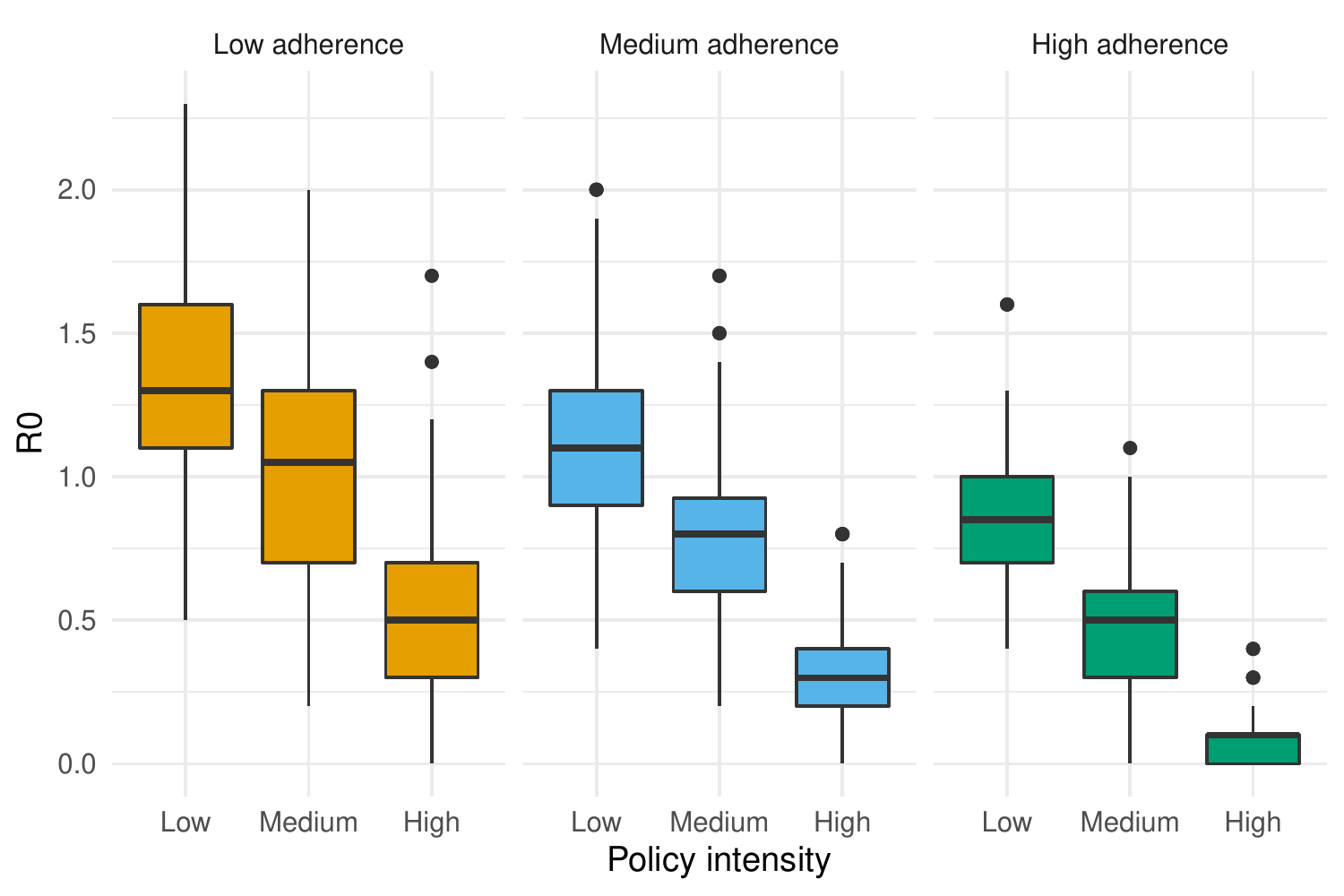}
    \caption{The values of $R_0$ for different policy and adherence intensity levels.} 
    \label{fig:R0_matrix}
\end{figure}

%\section{Network figures}

\section{Tables}

\begin{table}[H]

\begin{tabular}{rlllllll}

& \textbf{Single} & \textbf{Double} & \textbf{Smls} & \textbf{Mds} & \textbf{Lrgs} & \textbf{Seats} & \textbf{Capacity} \\
\cmidrule[\heavyrulewidth]{2-8}
Small Dorm       & 5       & 5       &        &         &        &       & 15       \\
Medium Dorm      & 15      & 15      &        &         &        &       & 45       \\
Large Dorm       & 25      & 25      &        &         &        &       & 75       \\
Small Clsrm  &         &         &        &         &        & 10    & 15       \\
Medium Clsrm &         &         &        &         &        & 15    & 20       \\
Large Clsrm  &         &         &        &         &        & 20    & 30       \\
%\hline
\cmidrule(lr){2-8}
Small Acad      &         &         & 3      & 0       & 0      & 30      & 45       \\
Medium Acad     &         &         & 2      & 3       & 0      &  65     & 90      \\
Large Acad    &         &         & 5      & 3       & 3      &    155   & 225      \\
%\hline
\cmidrule(lr){2-8}
Dorm Bldgs       &         &         & 25     & 10      & 10     &       & 1575     \\
STEM Bldgs &         &         & 2      & 2       & 3      & 655   & 945      \\
Humanities Bldgs       &         &         & 1      & 2       & 1      & 315   & 450     \\
Arts Bldgs       &         &         & 2      & 1       & 1      & 280   & 405    \\
\cmidrule[\heavyrulewidth]{2-8}
\end{tabular}
%\vspace{.25 cm}
\caption{At the top, counts for the number of single and double dorm rooms, the number of seats in classrooms. In the middle, the number of classrooms in each type of building. On the bottom, the number of each type of building.} \label{tbl:buildings}	
\end{table}

\begin{table}[H]
\begin{tabular}{l @{\hskip 0.25in}ccc@{\hskip 0.35in}  cc@{\hskip 0.35in} cc}
%\hline
& \multicolumn{3}{l}{\textbf{On-Campus}} & \multicolumn{2}{l}{\hspace{-.15in}\textbf{Off-Campus}} & \multicolumn{2}{l}{\textbf{Faculty}} \\
%& & On &  & & Off  & Fac &   \\
  & $A$ & $B$ & $W$  & $A$ & $B$  & $A$ &$B$ \\
  %\toprule
\cmidrule[\heavyrulewidth]{2-8}
8  & D &  D & D & & & &\\
9  & DH & D & DH  & OC & OC  & OC & OC\\
10 & $C_1$ & DH &D & $C_1$ & L &O &O \\
11 & $C_1$ & S & L & $C_1$ & S & O & O \\
12 & DH & $C_4$ & S  & DH & $C_4$ & DH & O  \\
13 & S  & $C_4$ & DH & L & $C_4$ & O  & DH \\
14 & $C_2$ & DH & S & $C_2$ & DH & $C_1$ & $C_2$ \\
15 & $C_2$ & G & G  & $C_2$ & G  & $C_1$ & $C_2$\\
16 & $C_3$ & D  & L & $C_3$ & L &O &O  \\
17 & $C_3$ &S &L & $C_3$ & S & O & O \\
18 &DH & D  & D & OC & OC & OC & OC \\
19 & L  & DH & DH  & & & &\\
20 & S  & D  & S   & & & &\\
21 & D  & D & S  & & & &\\
22 &D &D &D  & & & &\\
%\hline
\cmidrule[\heavyrulewidth]{2-8}
\end{tabular}%
\quad%
\begin{tabular}{ll}
    \multicolumn{2}{c}{\textbf{Key}} \\
    \toprule
   D  & Dorm \\
   DH & Dining Hall \\
   $C_i$ & $i$th class \\
   S & Social Space \\
   L & Library \\
   G & Gym \\
   OC & Off Campus \\
   O & Office \\
   %\bottomrule
\end{tabular}%
%\vspace{.25 cm}
\caption{Sample schedules for an on-campus student, an off-campus student, and a faculty member. Each row is the time of day.
%D = Dorm, DH = Dining Hall, $C_i$ = $i$th class, S = Social Space, L = Library, G = Gym, OC = Off-campus, O = Office.
}
\label{tbl:schedule} 
\end{table}

\begin{table}[H]
\begin{tabular}{l@{\hskip 0.3in}  l l @{\hskip 0.3in} l l}
%\hline
	 & \, \, \,  \textbf{Core} & & \, \, \, \textbf{Leaf} & \\
	\textbf{Space} & $C_v$ & $r_v$ & $C_v$ & $r_v$ \\
%\hline
\toprule
	Transit Space & $100n$ & 1 &  & \\
	Dining Hall & $650$  & 1 & $100$  & 2 \\
	Faculty Dining Leaf &   &  & $20$  & 2 \\
	Library  & $10 \cdot 300$ & 1 & $50$   & 2 \\
	Gym & $10 \cdot 60$ & 3 & $10$  & 3 \\
	STEM Office & $10\cdot6\cdot50$ & 1 & 50  & 2 \\
	Hum/Art Office & $10 \cdot 6 \cdot 25$ & 1 & 20 & 2 \\
	Social Space &  &   & $10$ & 3 \\
	Large Gatherings & $40\lceil x / 40 \rceil$ & $3$ &  &  \\
	Small Acad  & $10\cdot 45$ & 1 & & \\	
	Medium Acad & $10\cdot90$ & 1  & &\\
	Large Acad  & $10\cdot225$ & 1  & &\\
	Small Clsrm &  &  & 15 & 2\\
	Medium Clsrm &  &  & 20 & 2\\
	Large Clsrm &  &  &  30 & 2 \\ 
	Single Dorm &  &  &1 & 3\\
	Double Dorm &  &  &2 & 3 \\
	Small Dorm  & $10 \cdot 15$ & 2 & $x$ & 3 \\
	Medium Dorm  & $10 \cdot 45$ & 2 & $x$ & 3 \\
	Large Dorm &  $10 \cdot 75$ & 2 & $x$ & 3  \\
%\hline
\bottomrule
\end{tabular} 
%\vspace{.25 cm}
\caption{The core and leaf capacity and risk multiplier for different buildings. The quantity $x$ is the number of people assigned to that space.}\label{tbl:capacaties} 
\end{table}

%\newpage 

\bibliographystyle{amsplain}
\bibliography{covid}

\end{document}